\input harvmac

\def \four{{\textstyle {1\ov 4}}}
\def \TM {type IIB$_{(1,9)}$\ }
\def \TE {type IIB$_{(0,10)}$\ }

\def \del {\partial}

\def \ha{{\textstyle{1\over 2}}}

\def \b {\beta}
\def \chi {\chi}

\def \p {\phi}
\def \m {\mu}
\def \n {\nu}

\def \t {\tau}

\def \C {{\cal C}}

\def \inv {^{-1}}
\def \ov {\over }
\def \four{{\textstyle{1\over 4}}}

\def \lr { \lref}
\def\np {{  Nucl. Phys. }}
\def \pl {{  Phys. Lett. }}
\def \mpl {{ Mod. Phys. Lett. }}
\def \prl {{  Phys. Rev. Lett. }}
\def \pr  {{ Phys. Rev. }}

\baselineskip8pt
\Title{
\vbox
{\baselineskip 6pt{\hbox{CERN-TH/96-359}}{\hbox
{Imperial/TP/96-97/13}}{\hbox{hep-th/9612164}} {\hbox{
  }}} }
{\vbox{\centerline { Type IIB instanton  
 }
\vskip4pt
 \centerline {as a wave in  twelve  dimensions                }
}}
\vskip -20 true pt
\medskip
\medskip
\centerline{   A.A. Tseytlin\footnote{$^{\star}$}{\baselineskip8pt
e-mail address: tseytlin@ic.ac.uk}\footnote{$^{\dagger}$}{\baselineskip8pt
Also at Lebedev  Physics
Institute, Moscow.} }

\smallskip\smallskip
\centerline {\it   Theory Division, CERN, CH-1211  Geneve 23,
Switzerland}
\smallskip\smallskip
\centerline {\it  and  }
\smallskip\smallskip
\centerline {\it Blackett Laboratory, 
Imperial College,  London,  SW7 2BZ, U.K. }

\bigskip\bigskip
\centerline {\bf Abstract}
\medskip
\baselineskip10pt
\noindent
0-brane of type IIA  string theory can be interpreted
as a dimensional reduction of a gravitational wave in 11 dimensions.
We observe that a similar interpretation  applies also to the 
D-instanton  background  of type IIB theory: 
it can be viewed as a reduction (along one spatial and one time-like
direction) of a wave in a 12-dimensional theory. The instanton charge
is thus related to a linear momentum in 12 dimensions.
This suggests that the instanton should play as important role in
type IIB theory as the 0-brane is 
supposed to play  in type IIA theory.

\Date {December 1996}
\noblackbox
\baselineskip 14pt plus 2pt minus 2pt

\lr\rust{J.G. Russo and A.A. Tseytlin, hep-th/9611047.}
\lr \johns {J.H.  Schwarz, \pl B360 (1995) 13,
hep-th/9508143.}

\lr\TT{A.A. Tseytlin, \mpl A11 (1996) 689,   hep-th/9601177.}
\lr \vafa{C. Vafa, \np B469 (1996) 403; hep-th/9602022.} 
\lr\town {P.K. Townsend, \pl B350 (1995) 184,
 hep-th/9501068; \pl B373 (1996) 68,  hep-th/9512062.
  }

\lr\pol {J. Polchinski, \prl 75 (1995) 4724, hep-th/9510017;  
 hep-th/9611050.} 
 
\lr \green {M.B. Green, \pl B266 (1992) 325; \pl B329 (1994) 435;
\pl B354 (1995) 271, hep-th/9504108; 
J. Polchinski, \pr D50 (1994) 6041, hep-th/9407031.} 

\lr \grg{ M.B. Green and M. Gutperle, hep-th/9612127.} 
\lr \ishi {N. Ishibashi, H. Kawai, Y. Kitazawa and 
A. Tsuchiya, hep-th/9612115. }

\lr \periw{ V. Periwal, hep-th/9611103.   } 
\lr \ggp { G.W. Gibbons, M.B.  Green
and M.J.  Perry, 
  Phys. Lett. B370 (1996) 37, hep-th/9511080.}
\lr \kleb{S.S. Gubser, A. Hashimoto, I.R. Klebanov and 
 J.M. Maldacena, 
 Nucl. Phys. B472 (1996) 231, hep-th/9601057.
} 
\lr\hull { C.M. Hull,
 Nucl. Phys. B468 (1996) 113, hep-th/951218.} 
\lr\vafa { C. Vafa, \np B469 (1996) 403, hep-th/9602022.} 
\lr\twel{M. Blencowe and M.J.  Duff, \np B310 (1988) 387;
H. Ooguri and C. Vafa, \np B361 (1991) 469;
\np B367 (1991) 83;
D. Kutasov and E. Martinec, \np B477 (1996) 652, hep-th/9602049;
hep-th/9612102; 
D. Kutasov, E. Martinec and M. O' Loughlin,
 \np B477 (1996) 675, hep-th/9603116;
A.A. Tseytlin, \np B469 (1996) 51, hep-th/9602064;
 D. Jatkar and K. Rama, \pl B388 (1996) 45, hep-th/9606009;  
  I. Bars, \pr D54 (1996) 5203, hep-th/9604139, hep-th/9607112,
 hep-th/9610074;
 S. Hewson and M.J. Perry, hep-th/9612008.}

\lr\ferr {S. Ferrara, R. Minasian and
             A. Sagnotti, Nucl. Phys. B474 (1996) 323,   
             hep-th/9604097. }
\lr \john { J.H. Schwarz,    \np B226 (1983)
269;   P.S. Howe and P.C. West, \np B238 (1984) 181.} 
\lr \berg  {E. Bergshoeff, C. Hull and
 T. Ort\' \i n, \np B451 (1995) 547, hep-th/9504081;
 E. Bergshoeff,  H.J. Boonstra  and T. Ort\'in, 
  Phys. Rev. D53 (1996) 7206, 
 hep-th/9508091.
 }

\lr \schw {J.H.  Schwarz, \pl B360 (1995) 
13 (E: B364 (1995) 252),
hep-th/9508143.}         
\lr \cosm  { B.R. Greene, A. Shapere, C.
Vafa and S.T. Yau, \np B337 (1990) 1.}  
\lr\banks{
T. Banks, W. Fischler, S.H. Shenker and L. Susskind, 
   hep-th/9610043.}

\lr\doug{M. Douglas, hep-th/9512077.} 
\lr\kab{M.R. Douglas, D. Kabat, P. Pouliot and  S.H. Shenker,
  hep-th/9608024. }
  \lr\park{J.-S. Park, hep-th/9612096.} 
\lr \wit{E. Witten, \np B460 (1996) 335, hep-th/9510135.}

Recently,  
it was   suggested   \refs{\town,\banks} that 
0-branes \pol\ 
may be considered as  basic building  blocks of M-theory. This is  
related to the fact that the 0-brane charge  can be interpreted
as a 11-dimensional momentum. At the level of the 
 classical solutions, 
the $D=10$ 0-brane background is just a dimensional reduction of a
gravitational  wave propagating in 11 dimensions, 
$ds^2_{11} = -dt^2 + dx_{11}^2 + {q\ov r^7} (dt-dx_{11})^2 +
dx_idx_i$. The fundamental nature of the 0-brane is also indicated 
 by the
fact that other extended  objects in the theory can be `constructed'
out of arrays  of 0-branes  by  duality transformations. 

It seems important  to understand the type IIB theory 
analog of this 
picture. The object of minimal dimensionality here is the
D-instanton  \refs{\green,\pol,\grg}. It is  related  
to D0-brane by formal  T-duality in the  time direction. This  is a hint  that 
the instanton should play  a central   role in 
(a 12-dimensional reformulation of)  type IIB theory. In fact, 
 the recent proposal of a  matrix model  behind type IIB theory
 \ishi\ which is  based on a large N limit of the
 zero-dimensional  reduction of  SU(N) 
  10-d super Yang-Mills theory   may    
 be interpreted in 
 this  way ($L= \tr([A_\m,A_\n]^2 + 2i \bar \psi \gamma^\m
 [A_\m,\psi])$ \ 
 is the leading term in the action for N  D-instantons 
  \refs{\wit,\grg}).\foot{In view
 of T-duality  
 between instantons and 0-branes, 
 it is not surprising  that this action can be formally 
  related \ishi\ 
 to the 0-brane action used in \banks. A  similar 
  action but in 12 dimensions  was suggested  in connection  with 
   the action of \banks\ in \periw.}

Below we shall provide  a new   evidence 
of  a  fundamental nature of the type IIB instanton, and, at the
same time, of an existence of a
12-dimensional  structure  behind type IIB theory: 
just like the type IIA 
0-brane corresponds to a gravitational wave in 11 dimensions, 
the type IIB instanton  is an `image' of  a
gravitational wave in 12 dimensions.
 In particular, 
the instanton charge  is this identified with 
  the  12-dimensional momentum. 
 
 Our discussion of the 12-dimensional interpretation of the
 type IIB  instanton  solution of \ggp\ (see also \kleb) 
will be in the spirit of  the F-theory 
  proposals  in \hull\ and,
especially, in  
\vafa.\foot{For other suggestions
about   12-dimensional origin of  type IIB theory see 
\twel.} There will be an  important new   point: we 
  will  need to 
consider the {\it euclidean} 
version of type IIB theory (with signature (0,10)), 
and thus will compactify 
the  $D=12$ theory with signature (1,11) on  
a 2-space of (1,1) signature.

While  a  hypothetical   12-dimensional theory which leads to type IIB 
theory upon dimensional reduction  can not be of the standard 
supergravity type, it may not be that different, assuming    certain 
additional constraints are imposed. The $SL(2,R)$ 
 structure of the  field equations \john\ or the action \berg\
 of type IIB theory  provide strong  hints  about its 
   12-dimensional counterpart. Let us first consider the
   \TM theory  with Minkowski signature. 
   The  12-dimensional action  should    contain
  at least  the Einstein  term, probably   
  supplemented with certain conditions on the $D=12$ metric. 
  We shall  adopt  the following ansatz 
for the    metric ($\m,\n=0,1,...,9;\ p,q= 1,2$)
\eqn\mett{
ds^2_{12} = ds^2_{10E}  + ds^2_{2} =
 g_{\m\n} (x) dx^\m dx^\n  + M_{pq}(x) dy^p dy^q 
\    }
$$=  g_{\m\n} (x) dx^\m dx^\n  
+  e^{-\p(x)} dy_1^2 + e^{\p(x)}  [dy_2 + C(x) dy_1]^2 \ , $$
where $g_{\m\n},\p$ and  $C$ are  the Einstein-frame metric, dilaton and
R-R scalar 
of  IIB theory. The metric $M$ of an internal  2-torus 
with the complex structure modulus  $\t$ is  
\eqn\maay{
 M_{pq} =  e^\p \pmatrix{ e^{-2\p} + C^2  &  C   \cr   C   & 1  \cr}
  \ , \ \ \ \ \tau= C + ie^{-\p} \ , \   \ \ \ \det M_{pq}=1 \ .  }
Note that in contrast to the  
 similar  relation between  the $D=11$ supergravity and type IIA
theory  metrics, i.e.  
 $ds^2_{11}  =  e^{-\p/6} ds^2_{10E}
+ e^{4\p/3} (dy  + A_\m dx^\m)^2$, 
the ansatz \mett\ 
is not of the most general type: the volume of the  internal
2-torus is assumed 
to be  non-dynamical  ($\det M=1$)  \vafa\ as there are only 
two scalars in type IIB action.\foot{This restriction  
may be related to some  extra   symmetry (conformal invariance?)
 of the $D=12$ theory.}
 Dimensional  reduction then gives 
\eqn\acc{
S= \int d^{12} x \sqrt{ g_{(12)}} R_{(12)} =
\int d^{10} x \sqrt g[
 R +  \four   \Tr (\del_\m M \del^\m  M^{-1}) ]}
 \eqn\accc{ = \int d^{10} x \sqrt g [R - \ha \t^{-2}_2 {|\del \tau|^2} ]
 = \int d^{10} x \sqrt g[R - \ha (\del \p)^2
 - \ha e^{2\p} (\del C)^2 ]  
  \ . }
The metric \mett\ and the action \acc\  are  covariant under the 
 $SL(2,R)$ transformations, explaining the corresponding symmetry  
of type IIB theory \refs{\hull,\vafa}. 
 Other terms in the bosonic part of the type IIB
action can be understood by making a bold assumption
 that the $D=12$
theory should  contain also the 3-rank and 4-rank antisymmetric tensors
$C_3$ and $C_4$.\foot{It may be more natural 
to assume  that the fundamental field of $D=12$ 
theory  is only $C_4$  (which  already has enough 
components)  while $C_3$ is related to it by some constraint.
That would also make it clear that the $D=12$ theory should contain
only the 3-brane and 5-brane  extended objects (see also below).}
Then the two  2-rank tensors of type IIB theory $B_{p}$ 
($p=1,2$)
appear as $C_{\m\n p}$  components of $C_3$
and, moreover, the natural kinetic term
 $F^2(C_3)$ reduces to  another important 
 $SL(2,R)$ covariant structure 
  in the type IIB action, $ M^{pq} dB_p dB_q$.
  The $D=12$ Chern-Simons coupling 
  $\int C_4 \wedge dC_3\wedge dC_3$
 produces the needed $\int C_4 \wedge dB^{(1)} \wedge dB^{(2)}$
term in IIB theory action  \ferr.
 Obviously, there should be other magical 
constraints  that  should (i) rule out various extra  terms which 
 appear from  $ \int [R_{(12)} - F^2(C_3) - F^2(C_4) -
C_4 \wedge dC_3\wedge dC_3 + ...]$  upon  
 direct  dimensional reduction, 
 (ii) imply self-duality of the field strength 
of $C_4$ in $D=10$ and, 
of course,  (iii) ensure the existence of supersymmetry.

Assuming the existence of  such  12-dimensional 
theory, it should be  possible to relate type IIB 
 p-brane solutions 
 to certain 12-dimensional field  configurations.
 It  is  natural to expect that the $D=12$ theory 
should have   3-brane and 5-brane solutions (which are 
`electro-magnetic' dual  in $D=12$). 
The $SL(2,Z)$ family of type IIB strings \schw\ 
then  may   appear as wrappings of the 12-dimensional 
 3-brane around the internal 2-torus.  
To understand  such  relation in detail one  first needs 
 to clarify
 the  structure of the antisymmetric tensor field couplings 
 in  the 
$D=12$ theory (in particular, the relative roles of $C_3$ and
$C_4$).\foot{Naive wrapping of 3-brane with $C_4$ charge does not 
seem  to give the charges of the 
 $B_p$-fields related to $C_3$ as discussed above.
One also needs to understand  
 how to connect the $SL(2,Z)$ family
of type IIB 5-branes to the $D=12$ 5-brane: a puzzle here 
is that the  internal 2-torus should not be part of the 
5-brane
(see also below).}

In what follows  we shall concentrate on purely gravitational 
$D=12$ backgrounds which do not depend  on  unknown  details of 
the  structure of the antisymmetric tensor sector. 
Like   the 0-brane and the  6-brane of type IIA theory which 
correspond to the   gravitational  solutions
in $D=11$ theory (plane wave and euclidean Taub-NUT or Kaluza-Klein
 monopole), 
the instanton and  the 7-brane of type IIB theory 
also have purely gravitational counterparts in $D=12$ theory.
The 7-brane  case   was already discussed in \vafa.
The solution corresponding to a collection of  $n$
  type IIB 7-branes 
\refs{\ggp,\cosm}  
 is given by \mett\  of the following  special form
  ($z=x_8 + ix_9$)  
\eqn\seven{
ds^2_{12} =  -dt^2  + dx_1^2 + ... +dx_7^2  
+  H^2(z,\bar z) dz d\bar z +   H\inv (z,\bar z)  |dy_2 +
\tau (z) dy_1|^2  \ , }
where $H = e^{-\p} =\tau_2 ,$ and  $j(\tau (z)) 
= P_n(z)/P_{n-1}(z)$.  
The  regular case of $n=24$   7-branes on a compact $(z,\bar z)$ 
2-space 
 in type IIB theory
can be  interpreted  \vafa\
as  a special K3 compactification \cosm\ of 
the 12-dimensional theory. 

Our aim  here is to give a similar interpretation
to the type IIB  D-instanton.
The instanton is a solution \ggp\ of the {\it euclidean}
type IIB theory 
(which has a  well-defined 
euclidean supersymmetry) 
  with the action  \accc\ 
where  $g_{\m\n}$ is 
 assumed to have  euclidean signature 
 and the scalar $C$ is   replaced by $i \C$. 
This rotation of $C$ has a $D=10$ explanation if type IIB theory
is defined in terms of the dual $F_9$ field strength \ggp.
At the same time,  it has also an alternative
 natural   $D=12$
explanation if  the {\it euclidean}\   \TE theory 
corresponds  to
a  compactification
of the same 12-dimensional  theory of the  signature (1,11)
 but now on a 2-space of  the  signature (1,1).
If  $g_{\m\n}$ in \mett\ is taken to be euclidean, 
the coordinate $y_1$   should become  time-like, 
$y_1=-it$.
To preserve the  reality of the metric \mett\ one should then  rotate 
$C
\to i\C$. The result is  
($y\equiv y_2$; \ $m,n=1,2,...,10$)  
\eqn\mink{
ds^2_{12} = 
  - e^{-\p(x)} dt^2 + e^{\p(x)} [dy + \C(x) dt]^2 
   + g_{mn} (x) dx^m dx^n  \ .  }
The \TE theory is then  obtained  
by dimensional reduction in the 
spatial direction $y$ and the {\it time-like}  direction $t$
(cf. \refs{\vafa,\twel}). 

One of the simplest examples of such a 
  gravitational background is a spherically symmetric 
pp-wave, 
\eqn\wave{
ds^2_{12} = 
  - dt^2 + dy^2 +  [H(x)-1]  (dt-dy)^2  + dx_m dx_m 
  }
  $$
  =   - H\inv  dt^2 +  H  [dy  +  (H\inv-1) dt]^2  + dx_m dx_m 
   \ ,
  \ \ \ \ \  \   H= 1 +  {q\ov x^8}  \ .   $$
It solves the vacuum Einstein equations 
 and should be supersymmetric (as is always 
the case in lower dimensions), provided
a supersymmetric $D=12$ theory can be defined.
Comparing \mink\  to  \wave\  we learn that 
the corresponding \TE background is exactly the instanton solution 
of \ggp\
\eqn\inis{
 e^\p =  H(x) \ , \ \ \ \ \  \C=    H\inv (x) - 1\ , \ \ \ 
\ \ \  ds^2_{10E} = dx_m dx_m  \ . }
We assumed that the fields have trivial asymptotic values
 $ \p_\infty=0,\ \C_\infty=0$, 
i.e. that the vacuum `2-torus' $T^{(1,1)}$ is trivial.
The analog  of the $SL(2,R)$ symmetry in \TE theory 
acts only on  the constant parameters $(g=e^{\p_\infty},\  \C_\infty)$
of the generic  solution. The constant $q$ is related 
to the instanton charge $Q_{-1} = 8 \omega_9 q= {2\ov 3} \pi^{5/2}
q$ \ggp, which    can now be interpreted as 
 a linear momentum carried by the wave in the 12-th dimension.
 As in the case of the  `0-brane charge -- 11-dimensional
 momentum' correspondence,
 this provides  another  reason for the  quantisation of $Q_{-1}$.   

Let us now discuss  some  implications of the 
above observation. Suppose that the metric \wave\
has an extra spatial isometry in one of the $x_m$ directions, e.g., in  
$x_{10}\equiv z$  (then  $H= 1  +  {q\ov x^7}$). The  reduction 
along $(t,y)$ then connects  it  to a \TE background produced 
by a periodic array of instantons in $z$ direction. 
If we  also assume that the 12-dimensional theory is somehow related
 to M-theory 
by a reduction in $z$ (more generally, 
 that a compactification of the (2,10)  theory
on $T^{(1,1)} \times S^1$  corresponds  to a compactification of  (1,10)
theory on $S^1\times S^1$, cf. \vafa) 
 then the resulting  11-dimensional
background is again a similar  plane wave.  Further reduction 
along $y$ leads to  the  0-brane solution of type IIA theory
with the following string-frame metric,  dilaton and  vector 
field:  
$ds^2 = H^{1/2}( - H\inv dt^2 + dx_k dx_k), \ e^{\p} 
=H^{3/4}, \ A_t = H\inv -1$.  
The latter background is related to the above type \TE  solution 
(the one which is `smeared' in $z$-direction) 
 by formal T-duality in $t$
($A_t\to \C$, etc.) 
and the identification of the dual $t$-coordinate with
$iz$ ($T$-duality in time direction transforms a real 
background   in IIA theory into a complex one in \TM  theory  but again 
 a  real  background  in \TE theory). 
The  consistency of this picture seems 
to suggest that like the $SL(2,Z)$ symmetry of type IIB theory, 
the $T$-duality
between   type IIB and type IIA theories may have a simple
origin in the 12-dimensional theory, being  related to 
 a coordinate transformation  interchanging $y$ and $z$ compactification
 directions.\foot{This is also  implied by  the fact that the $D=9$ 
 theory  now appears as 
 a reduction of  the $D=12$ theory on the $(t,z,y)$ space  of signature
 (1,2), thus   suggesting an 
  explanation for  the $D=9$  U-duality symmetry $SL(2,R) \times
  SO(1,1)$ \berg.} 

Finitely boosting  0-brane in one extra isometric direction $x_9$
(this corresponds to a wave along  generic cycle of 2-torus
$(x_{11},x_9)$
in $D=11$ theory \rust) 
and doing $T$-duality in $x_9$ (i.e. performing $O(2,2)$ duality 
on the  0-brane background) leads to the $SL(2,Z)$  family 
of strings \schw\ in \TM theory. Its counterpart in \TE theory, 
which may be
interpreted as a mixture of a `smeared' instanton and a string, should 
correspond to a reduction of a  3-brane  configuration  
(with  non-trivial  antisymmetric tensor background) 
in $(1,11)$ theory.

The T-duality between the `smeared' type IIB instanton 
and the 0-brane implies also the existence of a  non-supersymmetric
(non-BPS) 
generalisation of the `smeared'  instanton (`black instanton').
 Applying 
 T-duality to the non-extremal 0-brane (which is a dimensional
reduction of the  $D=11$ Schwarzschild background   finitely boosted
in an additional isometric direction, with the extremal case 
corresponding to the infinite boost limit) we find
the following \TE  solution  (cf. \inis)
\eqn\insno{
 e^\p = \hat  H(x) \ , \ \ \  \C= \coth \b\  
 [   \hat H \inv (x) - 1]\ , \ \ \ 
  ds^2_{10E} = f\inv (r) (dz^2 + dr^2) + r^2 d\Omega_8^2 \ , }
$$ f= 1 - {\mu\ov r^7} \ , \ \ \ \ \ 
\hat H= 1 + { \hat q \ov r^7} \ , \ \  \ \ \  \hat q = \m\ {\rm sinh}^2\b =
q\ \tanh\b \ ,  $$
where $\m$ is the non-extremality and $\b$ is the boost parameter.
For zero charge $q=0$ this metric is  T-dual  to the $D=10$ 
Schwarzschild metric  in the euclidean time direction ($t=iz$).

The  relation between the instanton charge 
and  the 12-momentum  suggests also an interpretation of 
instantons bound to euclidean 3-brane world-volume 
 in \TE 
theory\foot{By considering a 3-brane probe in 
the D-instanton background one finds that there is no non-trivial 
potential term in the 3-brane  action.
The  1/4 supersymmetric solution of \TE theory 
corresponding to a 
combination of a euclidean 
3-brane and an instanton 
 is easy to construct explicitly;  for example,   
the string-frame  metric is 
$ds^2_{10}= H^{1/2}_{-1}  H^{1/2}_3 
( H_3^{-1}  dx_k dx_k + dx_idx_i ), $
where $k=1,...,4$,\ $i=5,...,10$, and  $H_{-1} $ and $H_3$ are the 
instanton and the 3-brane harmonic functions depending on $x_i$.
An alternative  approach is to  assume that  
  the instanton charge  is generated 
   by the gauge field instanton of the 
   3-brane world-volume theory  
due to the presence of the $\int  \C  F\wedge F$  coupling  
in the 3-brane action \doug\ (this may  have a relation to a discussion
of D-instantons in \park).} as corresponding to  3-branes boosted  
in the twelfth
dimension. This is similar to the 11-dimensional 
interpretation of  analogous 
0-brane bound states in type IIA theory \refs{\kab,\rust}.
 
Finally, let us note that the above discussion 
of  \TE theory based on compactification
of $D=12$ theory  in one spatial and one time-like direction suggests
that a similar interpretation should be  possible also for \TM theory:
one is just to assume that the metric  $g_{mn}$  in \mink\ has  Minkowski
signature while  still reducing  in the $(t,y)$ directions.
The $D=12$ theory then has the $(2,10)$ signature 
as in some of the  proposals in
\refs{\vafa,\twel}.   Its  3-brane solution will then have to have 
(2,2)
world-volume signature (to be   related to type IIB strings upon 
compactification in $(t,y)$)
 while the 5-brane  may  still have the usual 
$(1,5)$ signature, possibly 
resolving the problem mentioned in footnote 5.

\medskip
I am very grateful to A. Schwimmer for a stimulating discussion.
 This work was partially supported by PPARC and 
 the European
Commission TMR programme ERBFMRX-CT96-0045. 

\vfill\eject
\listrefs
\end